# Enhancement of Photovoltaic Current Generation through Dark States in Donor-Acceptor Pairs of Tungsten-based Transition Metal Di-Chalcogenides (TMDCs)

Sayan Roy[1], Zixuan Hu[2,3], Sabre Kais[2] & Peter Bermel[1]

1. Department of Electrical and Computer Engineering and Birck Nanotechnology Center, Purdue University, West Lafayette, IN 47907, United States.
2. Department of Chemistry, Department of Physics, and Birck Nanotechnology Center, Purdue University, West Lafayette, IN 47907, United States.
3. Qatar Environment and Energy Research Institute, College of Science and Engineering, HBKU, Doha, Qatar.

Contact Information:
   Sayan Roy:      roy1@purdue.edu
   Zixuan Hu:      hu501@purdue.edu
   Peter Bermel:   pbermel@purdue.edu
   Sabre Kais:     kais@purdue.edu

**ABSTRACT**

As several photovoltaic materials experimentally approach the Shockley-Queisser limit, there has been a growing interest in unconventional materials and approaches with the potential to cross this efficiency barrier. One such candidate is dark state protection induced by the dipole-dipole interaction between molecular excited states. This phenomenon has been shown to significantly reduce carrier recombination rate and enhance photon-to-current conversion, in elementary models consisting of few interacting chromophore centers. Atomically thin 2D transition metal di-chalcogenides (TMDCs) have shown great potential for use as ultra-thin photovoltaic materials in solar cells due to their favorable photon absorption and electronic transport properties. TMDC alloys exhibit tunable direct bandgaps and significant dipole moments. In this work, we introduce the dark state protection mechanism to a TMDC based photovoltaic system with pure tungsten diselenide ($WSe_2$) as the acceptor material and the TMDC alloy tungsten sulfo-selenide (WSeS) as the donor material. Our numerical model demonstrates the first application of the dark state protection mechanism to a photovoltaic material with a photon current enhancement of up to 35% and an ideal photon-to-current efficiency exceeding the Shockley-Queisser limit.

## 1. Introduction

Since its discovery in the 1960s, the Shockley-Queisser (S-Q) efficiency limit has generally been viewed as a fundamental limit on the performance of conventional photovoltaic devices [1,2], because the detailed balance principle defines a non-trivial loss associated with the radiative recombination process. After decades of progress, recent work has brought certain high-performance photovoltaics made from multiple materials within several percent of this limit [3]. Solar cells have been shown to exhibit strong internal and external luminescence as they approach the S-Q limit, which limits their maximum efficiencies [4]. Concentrators are the most straightforward approach to increase the S-Q efficiency, but generally require operating at high temperatures, or with large cooling structures, and have not been widely adopted commercially [5]. Solar cells composed of nanophotonic structures have also been widely studied to help extend the S-Q limit (e.g., by restricting the range of incident angles for increased open circuit voltage), but current experimental devices have significant limitations in achieving higher efficiencies and stability [6,7,8]. As a result of these constraints, there has been a growing interest in many unconventional materials and approaches with the potential to break the S-Q limit [9,10]. Arguably, the most successful approach is multijunction photovoltaics, where the solar spectrum is split by stacking materials with different bandgaps [11]. However, the cost of publicly-known fabrication approaches can be orders of magnitude higher than single junctions [12]. In a related approach, multiple spectrum solar cells transform the broad solar spectrum to a narrow range of photon energies, such as in thermophotovoltaics [13]; however, a major challenge is in obtaining higher efficiency devices. Multiple absorption is a mechanism where a single, high-energy photon generates multiple electron-hole pairs for higher efficiencies [14], but there is significant difficulty in the subsequent transport and collection of carriers. Another approach to obtain increased incident energy absorption is hot carrier extraction [15], but a major challenge is fabricating a device to efficiently extract this excess energy. It is also possible to obtain AC solar cells by treating the incident photons as electromagnetic waves, as demonstrated in optical rectennas [16], but scaling such devices to optical frequencies is a major technical challenge. Another approach uses multiple energy levels for demonstrable increases in photon absorption, such as intermediate band solar cells [17] and quantum well solar cells [18]. In this work, we present another approach based on multiple energy levels in our absorber. While most prior techniques use these levels to increase photon absorption, our approach does not expect increased absorption. Instead, our approach uses the newly-formed energy levels to greatly reduce carrier recombination for more efficient carrier extraction. This allows for higher photon-to-current conversion with previously demonstrated levels of photon absorption.

As inspired by the Shockley-Queisser efficiency limit, a natural direction of designing high efficiency photovoltaic systems is to reduce the carrier recombination rate. To this end, recent studies [19-23] investigated the possibility of using quantum effects in chromophore complexes to improve solar cell performance [24,25]. Optically dark states created by dipole-dipole interaction between molecular excited states can reduce radiative recombination in exciton transfer, effectively increasing the photocell efficiency [19-23,26]. However, these studies are limited to elementary models consisting of few interacting chromophore centers, and the dark state protection mechanism has not been applied to a specific material-based photovoltaic model. In this work, we use a photovoltaic model based on transition metal di-chalcogenide (TMDC) materials to demonstrate the dark state protection mechanism's ability to enhance the photocurrent, thus getting closer to an actual material design to overcome the Shockley-Queisser limit.

Transition metal di-chalcogenides (TMDCs) have gathered an increasing amount of interest recently for photovoltaic (PV) applications [27-29] because of their promising electronic and optical properties. It is possible to obtain alloys of TMDCs by altering their composition to contain more than one kind of chalcogen atoms [30], leading to the formation of hybrid TMDCs with tunable electronic and optical properties. Here we look at tungsten-based TMDC alloys containing sulfur and selenium, whose properties are intermediate of tungsten dilsulfide ($WS_2$) and tungsten diselenide ($WSe_2$) with tunable direct bandgaps dependent on the sulfur and selenium concentrations.

We have modelled a solar cell composed of pure tungsten diselenide ($WSe_2$) as the acceptor material and a TMDC alloy, tungsten sulfo-selenide (WSeS), as the donor material, in analogy with a heterojunction solar cell. For the donor, two layers of WSeS are placed on top of each other and the dipole-dipole interaction between the two layers split the conduction band into a bright band and a dark band. The dark band can then enable the dark state protection mechanism and enhance the photocurrent generated. WSeS has a large permanent dipole moment, whereas $WSe_2$ does not have any permanent dipole moment, thus the donor-acceptor coupling is a dipole-induced dipole interaction.

The electronic properties of $WSe_2$ and WSeS were studied using density functional theory (DFT) calculations, from which the band structures and corresponding bandgaps, dipole moments and coupling energies were obtained. The exciton transfer dynamics is calculated by a model based on Pauli master equations [19,20] and the photocurrent is calculated by the standard model for photocells [21,24,25]. Our numerical model demonstrates the first application of the dark state protection mechanism to a material based photovoltaic system with a photon current enhancement of up to 35% and an ideal energy conversion efficiency exceeding the Shockley-Queisser limit.

## 2. Results and Discussion

*2.1 Band Structure of Tungsten-based Pure TMDCs and Alloys.* We investigated alloys of TMDC alloys by altering the composition to contain more than one chalcogen atom, leading to hybrid TMDCs with tunable properties. We calculated the band structures of both pure and hybrid TMDCs by density functional theory (DFT) calculations. The bandgaps of pure $WS_2$ and $WSe_2$ are 2.15 eV and 1.67 eV, respectively. For a tungsten-based TMDC alloy containing both S and Se atoms, the bandgap is an intermediate value depending on the chalcogen composition. The bandgaps of the alloys show an approximately linear decrease with increasing Se concentration from $WS_2$ (2.15 eV) to $WSe_2$ (1.67 eV) as shown in Fig. 1. All the hybrid alloys are found to have a direct bandgap, similar to $WS_2$ and $WSe_2$.

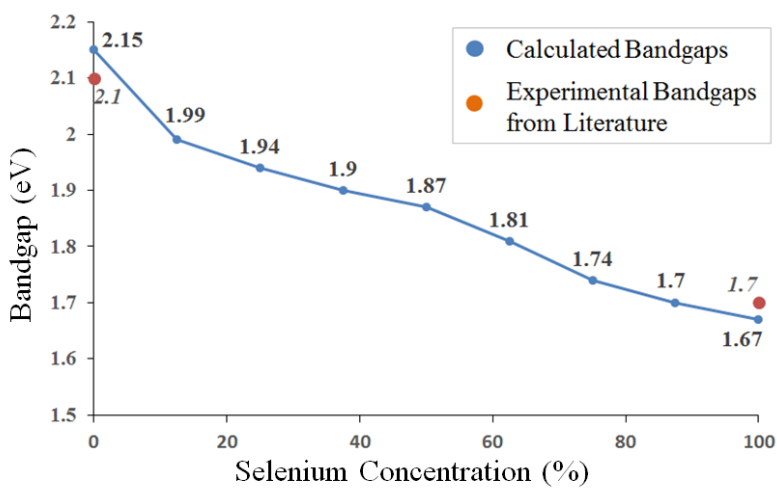

**Fig. 1**. Variation of bandgap in a tungsten-based TMDC alloy as a function of selenium concentration and comparison to experimental bandgaps from existing literature [31,32].

*2.2 Dipole Moment of Hybrid TMDCs.* The dipole moments of $WS_2$ and $WSe_2$ are zero, since they have same chalcogen atoms located symmetrically on both sides of the tungsten atoms. Starting from pure $WS_2$, the dipole moment was found to increase linearly with increasing Se concentration from 0% to 50%. For Se concentrations above 50%, the dipole moment decreased with further increase of Se concentration due to reappearance of symmetric atomic positions. Among the alloys, tungsten sulfo-selenide (WSeS) was found to have the highest dipole moment (0.27 D per molecule) since it has the most asymmetrical structure with only S atoms on one side of W atoms, and only Se atoms on the other side of W atoms.

*2.3 Donor-Acceptor Photovoltaic Model.* We use a donor-acceptor photovoltaic model composed of TMDC materials (Fig. 2). The material structure model is shown in Fig. 2(A): WSeS with a bandgap of 1.87 eV was chosen as the donor in our system; WSe$_2$ with a bandgap of 1.67 eV was chosen as the acceptor, since its lower bandgap favors the band alignment of our model. The donor consists of two layers of WSeS arranged on top of each other. There is strong dipole-dipole coupling between the two donor layers because of the large permanent dipole moment in each layer, and consequently the conduction band splits into a bright band and a dark band with an 18 meV energy gap. The induced polarization in the acceptor layer was calculated with the electric field from the donor layers (SI 3). The donor-acceptor coupling energy ($\gamma_C$) was calculated from the dipole-induced dipole interaction energy between the WSeS layers and the WSe$_2$ layer. For the equilibrium donor-acceptor spacing of 3 Å [33], the coupling energy is 515 μeV, which decreases with increasing donor-acceptor separation to a value of around 50 μeV for 10 Å separation (see the Supplementary Information SI 3). Radiative lifetimes of excitons in TMDCs are around 5-17 ps [34,35], corresponding to a recombination rate ($\gamma_R$) of 40-130 μeV. In our model, $\gamma_R$ is taken as 100 μeV. The exciton generation rate in our model is simulated by a hot photon bath at 5800K.

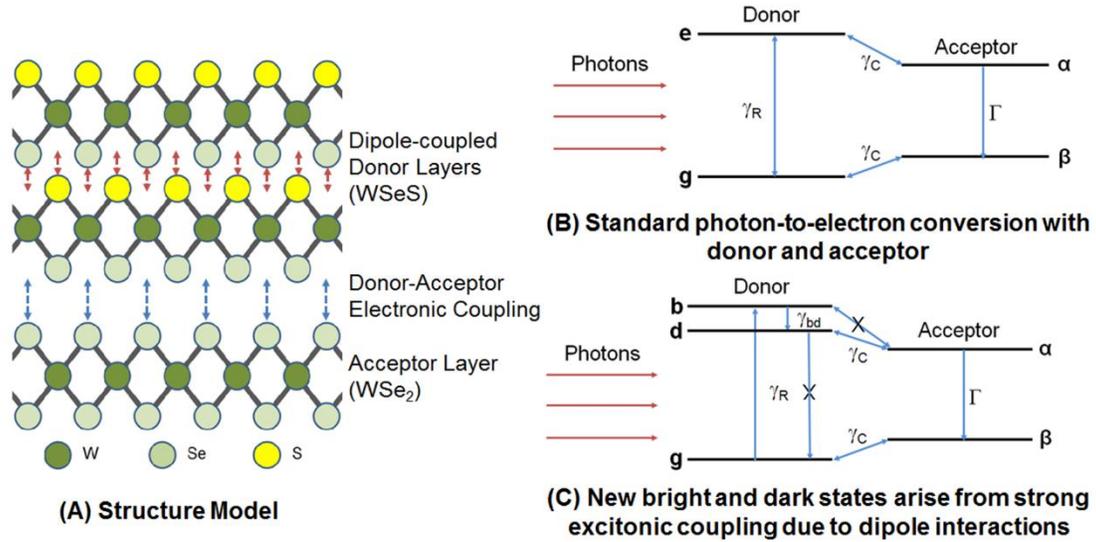

**Fig. 2.** Donor-Acceptor Photovoltaic Model

The energy diagrams and photovoltaic dynamical models are shown in Fig. 2(B) and Fig. 2(C). Fig. 2(B) shows the standard donor-acceptor model without dark state protection and Fig. 2(C) shows the improved model with dark state protection. In both models the incoming photons generate excitations in the donor with the $\gamma_R$ process, then transfer to the acceptor with the $\gamma_C$ process and convert to current with the $\Gamma$ process. The $\gamma_R$ process is reversible, leading to the radiative recombination loss. The new additional feature of our model is the formation of new optically excitable states through strong excitonic coupling between the donor layers, as shown in Fig. 2(C). This leads to a splitting of the conduction band in the donor with the formation of a bright state $b$ and a dark state $d$. Virtually all photon absorption and emission take place through the bright state. In this model, the initial absorption of photons leads to excitation on the bright state. The thermal relaxation process $\gamma_{bd}$ brings the excitations from the bright state $b$ into the dark state $d$ where radiative recombination is forbidden. Since radiative recombination cannot occur through $d$, photon re-emission is suppressed and fewer excitons are lost through recombination before getting transferred to the acceptor. The excitation is then transferred from

$d$ to $\alpha$. The photocurrent is determined by the rate $\Gamma$ from $\alpha$ to $\beta$. The dark state protection from recombination leads to an increase in the number of excitons available in the acceptor, and hence, gives a higher photocurrent in the solar cell. The overall dynamics is represented by Pauli master equations whose details can be found in the Supplementary Information (SI 4). Our results show that there is significant enhancement of the output photocurrent for the model in Fig. 2(C) compared to the one in Fig. 2(B) due to dark state protection. The current enhancement for different values of donor-acceptor separations and coupling energies are summarized in Fig. 3.

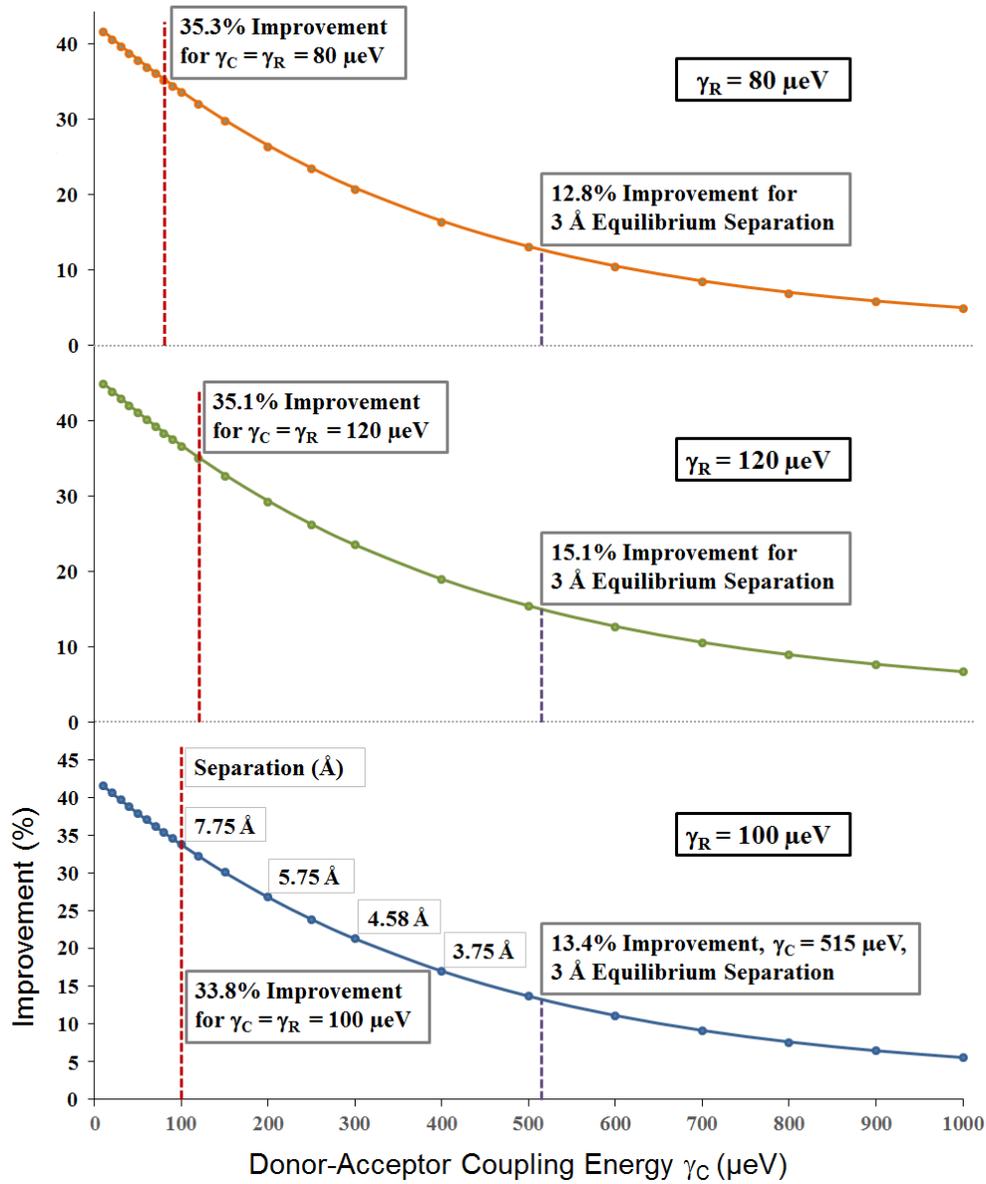

**Fig. 3.** Improvement of output current due to dark state protection for various donor-acceptor coupling energies for $\gamma_R = 100$ µeV, and comparative performance for different values of $\gamma_R = 80$ and 120 µeV (Data taken from [36]).

For the equilibrium separation of 3 Å and corresponding coupling energy of 515 µeV, an enhancement of 13.4% in the output photocurrent was obtained. The current of the system without dark state protection is 4.9 mA/cm², while introducing dark state protection increases the current to 5.56 mA/cm². If we look at the donor-acceptor separation scenario where the values of radiative recombination rate ($\gamma_R$) and donor-acceptor coupling energy ($\gamma_C$) are comparable to each other, we obtain a larger current enhancement of around 35% with the output photocurrent jumping from 0.99 mA/cm² to 1.34 mA/cm² due to dark state protection (for $\gamma_C$ = 80 µeV). The increase in current is the result of the suppression of radiative recombination in the donor material where the photon absorption and exciton generation take place. Fig. 3 shows that for systems where the donor-acceptor coupling energy is small compared to the rate of radiative recombination, we can obtain a greater percentage enhancement of the photocurrent by dark state protection. This is due to the competition between the donor-acceptor transfer process $\gamma_C$ and the recombination process $\gamma_R$. When the donor-acceptor transfer rate is small compared to the radiative recombination rate, significant loss can occur through the recombination process unless it is suppressed by dark state protection. By implementing dipole-dipole interactions and dark state protection in our donor-acceptor model (WSeS and WSe$_2$), a current enhancement of up to 35% can be obtained.

*2.4 Maximum Efficiency of Photovoltaic Model with Dark State Protection.* The maximum possible efficiency of a donor-acceptor solar cell model with dark state protection was calculated (Fig. 4) for the AM1.5G spectrum [37] and compared with the well-known Shockley-Queisser limit for single junction solar cells [1,2]. It can be seen that the efficiency limit of our model is significantly higher than the Shockley-Queisser limit over the entire bandgap range of 0.5-3 eV.

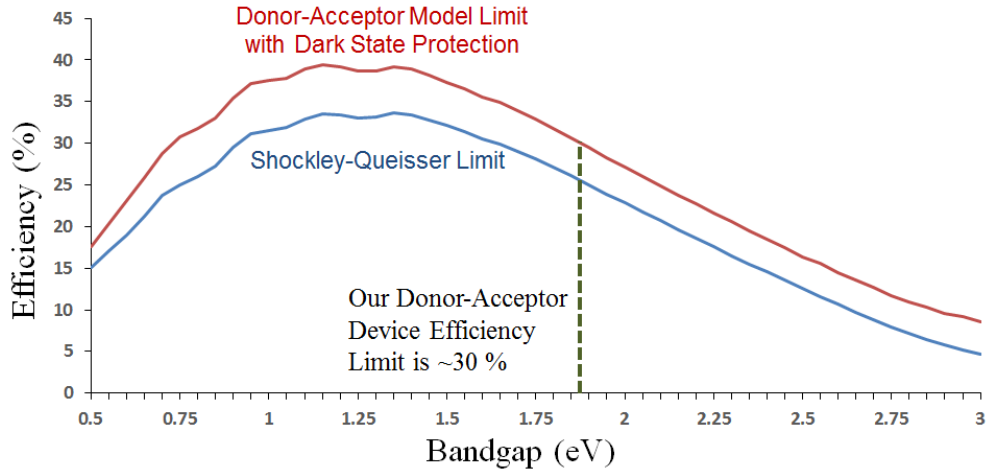

**Fig. 4.** Efficiency limit of a donor-acceptor photovoltaic system limit with dark state protection.

Our efficiency limit calculation is an idealistic approach to evaluate the efficiency of a donor-acceptor model with perfect dark-state protection. It is assumed that all the incident photons with energies above the donor bandgap are absorbed and correspondingly excitons are generated, there is no radiative recombination in the donor (dark state protection is perfect), and there is no loss of excitons during the $\Gamma$ process in the acceptor. We have accounted for carrier thermalization losses during the relaxation to the band edges in the donor and acceptor. For our donor material with bandgap 1.87 eV, we can get a maximum efficiency of around 30% using the donor-acceptor photovoltaic model with dark state protection under ideal conditions, compared to a maximum efficiency of around 26% as per the Shockley-Queisser limit for a semiconductor of the same bandgap 1.87 eV.

The efficiency limit calculation is a purely theoretical estimation as it doesn't account for many non-idealities; dark state protection might not be perfect at room temperature or with a conduction band splitting of less than 100 meV. A major challenge is obtaining 100% photon absorption in a only a few layers of the donor material. Also, there could be significant carrier recombination if the donor-acceptor coupling energy and the corresponding transfer rate is low, compared to the generation rate of excitons in the donor. On the other hand, if we can extract the hot carriers with energies above the band edge in the acceptor, it will lead to lower thermalization losses and allow for even higher efficiencies. Although not outright conclusive, the efficiency limit shown in Fig. 4 gives an insight into the possible improvements that can be obtained with incorporating the dark state protection mechanism into a TMDC-based photovoltaic system.

## 3. Conclusion

We have demonstrated a TMDC-based donor-acceptor photovoltaic model, where the dark state protection mechanism is used to reduce carrier recombination and enhance photon-to-electron conversion, leading to significantly higher current output. Transition metal di-chalcogenides (TMDCs) have already been shown to have great potential as ultra-thin photovoltaic materials in solar cells. In this work, we have explored and modeled a heterojunction-like solar cell composed of tungsten diselenide ($WSe_2$) as the acceptor material and tungsten sulfo-selenide (WSeS) as the donor material. The dipole-dipole coupling between the two layers of the donor material splits the conduction band and enables the dark state protection of excitation from recombination, achieving a photocurrent enhancement as high as 35% over the standard model without dark state protection. The enhancement is more significant when the donor-acceptor transfer rate is comparable to or smaller than the recombination rate. Although a preliminary estimate, the efficiency limit of such a model has been calculated to potentially overcome the Shockley-Queisser limit if all photons above the bandgap energy are absorbed with perfect dark state protection, and there is no loss during carrier collection in the acceptor. This opens up possibilities for exploring new materials and device architectures for ultra-thin, ultra-efficient photovoltaic devices.

# Methods

*4.1 Material Characterization.* The band structure and corresponding bandgap, and dipole moments of pure and hybrid TMDCs were obtained by density functional theory (DFT) calculations together with GW corrections using Quantum ESPRESSO [38-41]. TMDC alloys were investigated by altering the composition to contain more than one chalcogen atom, leading to hybrid TMDCs with tunable properties and significant dipole moments. More details about the calculations and modelling are described in the Supplementary Information (SI 1 and 2).

*4.2 Donor-Acceptor Exciton Transfer Dynamics.* The exciton transfer dynamics, including exciton generation from photons and corresponding output current, is represented by Pauli master equations. The details are explained in the Supplementary Information (SI 4).

*4.3 Maximum Efficiency with Dark State Protection.* The efficiency limit of a donor-acceptor solar cell model was calculated using the carrier and energy dynamics, and incident radiation in the AM1.5G spectrum [37]. For a particular bandgap of the donor, it is assumed that all the incident photons with energies higher than the bandgap are absorbed by the donor. The excited carriers in the donor are taken as relaxed to the band edge (dark state) for current extraction and the corresponding thermalization loss is taken into account. There is additional thermalization loss during the excitation transfer from donor to acceptor since the acceptor has a lower bandgap. The output energy obtained from each carrier is the bandgap of the acceptor where the excitons decay from the excited state to the ground state, accounting for the Carnot loss [2]. The range of donor material bandgaps was taken as 0.5-3 eV, with the acceptor bandgap taken to be 0.2 eV lower than that of the donor.

$$\eta = \frac{P_{IN}}{P_{OUT}} \times 100\%$$

where,

$\eta$ – Efficiency (%)

$P_{IN}$ – Total input power from AM 1.5G spectrum,

$P_{OUT}$ – Output power obtained from across the acceptor material bandgap.

**Acknowledgements**

The authors would like to thank Muhammad Ashraf Alam, Mark Lundstrom and Sangchul Oh for insightful discussions. Funding was provided by CAREER: Thermophotonics for Efficient Harvesting of Waste Heat as Electricity (NSF Award EEC-1454315); the Network of Computational Nanotechnology (NSF Award EEC-0228390); the United States Office of Naval Research (ONR) under Award No. N00014-15-1-2833; the U.S. Department of Energy, Office of Basic Energy Sciences under award number DE-SC0019215; and the Qatar National Research Fund exceptional Grant NPRPX-107-1-027.

# SUPPLEMENTARY INFORMATION

**Enhancement of Photovoltaic Current Generation through Dark States in Donor-Acceptor Pairs of Tungsten-based Transition Metal Di-Chalcogenides (TMDCs)**

Sayan Roy[1], Zixuan Hu[2,3], Sabre Kais[2] & Peter Bermel[1]

1. Department of Electrical and Computer Engineering and Birck Nanotechnology Center, Purdue University, West Lafayette, IN 47907, United States.
2. Department of Chemistry, Department of Physics, and Birck Nanotechnology Center, Purdue University, West Lafayette, IN 47907, United States.
3. Qatar Environment and Energy Research Institute, College of Science and Engineering, HBKU, Doha, Qatar.

**SI 1: Band Structure Calculation of Tungsten-based Pure TMDCs and Alloys**

TMDCs have a trigonal prismatic structure where each layer consists of a plane of hexagonally arranged transition metal atoms sandwiched between two planes of hexagonally arranged chalcogen atoms [1]. Density Functional Theory (DFT) was used to obtain the preliminary band structures of pure $WS_2$ and $WSe_2$. DFT is a powerful quantum mechanical method to obtain band structure of a material using many-body perturbation theory [2]. The DFT calculations were performed using Quantum ESPRESSO [3,4,5]. Quantum ESPRESSO is a quantum mechanical modelling tool to simulate many-body systems. DFT calculations require pseudopotential files which simulate individual atoms from the periodic table. GGA pseudopotentials (which add gradient correction to the more basic LDA ones), as described by the Perdew-Burke-Ernzerhof (PBE) scheme [6], were used to simulate the W and S atoms. The Brillouin zone was sampled according to the scheme proposed by Monkhorst-Pack [7] with a high-density in-plane $20\times20\times1$ k-point grid with 400 k-points. Since it is a thin 2D structure, there is no electronic dispersion in the vertical out-of-plane direction. The cutoff energy used was 30 Ry. The band structure obtained from DFT gives accurate information about the shape of the energy bands (such as VBM and CBM), but the bandgap is significantly underestimated. This problem was solved by formulating the electronic band structure using the GW approximation together with the DFT results; it involves the expansion of the self-energy in terms of the single particle Green's function G and the screened Coulomb interaction W to model many body systems [8,9,10]. The simulations were carried using the GWL package in Quantum ESPRESSO. The bandgap energy correction value obtained from the GW approximation was added to the DFT bands to obtain the accurate band structure of bulk TMDCs; however, for a TMDC monolayer, it is necessary to include the effect of the exciton binding energy [11]. The bandgaps of monolayer TMDCs obtained by DFT and GW calculations are higher than experimental results. This overestimation is due to the large excitonic effect in a two-dimensional system. When the energy correction due to the excitonic effect is subtracted from the DFT+GW bands, we get very accurate band structure and bandgap that match closely with experimental results [12,13,14]. All the calculations and simulations were carried out assuming a temperature of 300K. The bandgap of pure tungsten disulfide ($WS_2$) was calculated to be 2.15 eV (direct). The bandgap of pure tungsten diselenide ($WSe_2$) was found to be 1.67 eV (direct), as shown in Fig. S1(A).

TMDC alloys were investigated by altering the composition to contain more than one chalcogen atom, leading to hybrid TMDCs with tunable properties. Pure TMDCs have no dipole moment. Introducing multiple chalcogen atoms breaks the structure symmetry and leads to significant dipole moments. The bandgaps of $WS_2$ and $WSe_2$ are 2.15 eV and 1.67 eV, respectively. For a tungsten-based TMDC alloy containing both S and Se atoms, the bandgap is an intermediate value depending on the chalcogen composition. A periodic supercell containing 4 tungsten atoms and 8 chalcogen atoms was modelled. The alloys were modelled by introducing both S and Se atoms in the supercell, in all possible combinations starting from pure $WS_2$ to pure $WSe_2$, to obtain a hybrid structure. A relaxation calculation was carried out for each of alloy to obtain the lowest energy and most stable structure. The band structure and corresponding bandgap of each alloy were obtained by Density Functional Theory (DFT) calculations together with GW corrections as explained earlier. The band structure of the intermediate alloy, tunsten sulfo-selenide (WSeS), with a bandgap of 1.87 eV is show in Fig. S1(B).

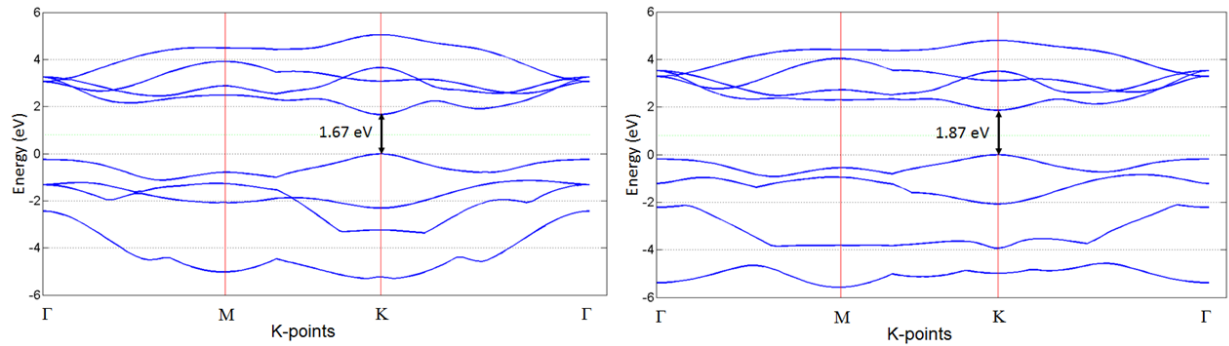

**(A)** Pure Tungsten Diselenide (WSe$_2$)    **(B)** Tungsten Sulfo-Selenide (WSeS)
**Fig. S1.** Band Structure of tungsten-based transition metal di-chalcogenides (TMDCs)

## SI 2: Dipole Moment Calculation

The DFT results from the band structure calculation were used to obtain the charge distribution in the pure and hybrid TMDCs, from which the dipole moment of each material was calculated. The calculations were performed using the post-processing operations in Quantum ESPRESSO [3]. The dipole moments of pure WS$_2$ and WSe$_2$ are zero, since they have same chalcogen atoms located symmetrically on both sides of the tungsten atoms. Starting from pure WS$_2$, the dipole moment was found to increase linearly with increasing Se concentration from 0% to 50%. For Se concentrations above 50%, the dipole moment decreased with further increase of Se concentration due to reappearance of symmetric atomic positions. Among the alloys, tungsten sulfo-selenide (WSeS) was found to have the highest dipole moment (0.27 D per molecule) since it has the most asymmetrical structure with only S atoms on one side of W atoms, and only Se atoms on the other side of W atoms. The charge distribution of the intermediate alloy, WSeS, through a vertical section is show in Fig. S2. The dipole moments of the hybrid and pure TMDCs are summarized in Table S1.

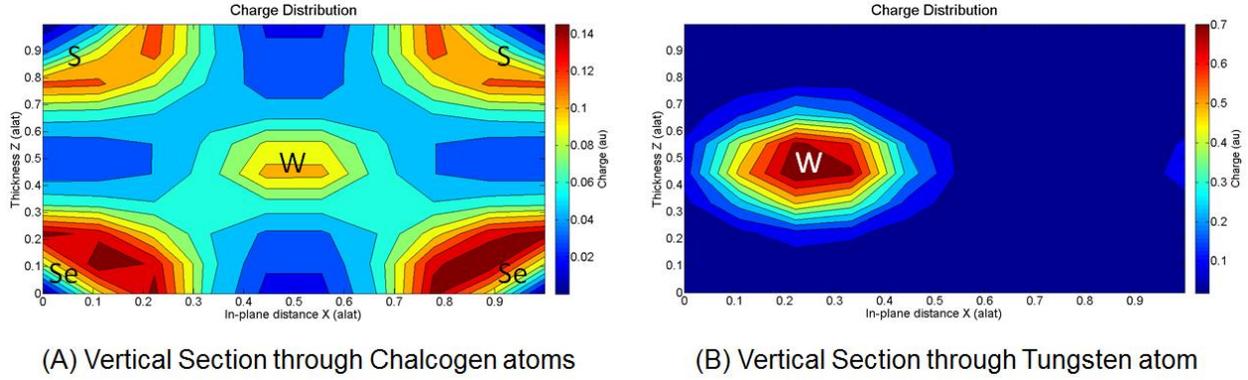

(A) Vertical Section through Chalcogen atoms    (B) Vertical Section through Tungsten atom

**Fig. S2.** Charge Distribution in WSeS

## Dipole Moments of TMDC Alloys [$WSe_{2X}S_{2(1-X)}$]

| Selenium Concentration (X) | Dipole Moment (D) |
|---|---|
| 0 | 0.0 |
| 0.125 | 0.07 |
| 0.25 | 0.14 |
| 0.375 | 0.21 |
| 0.5 | 0.27 |
| 0.625 | 0.21 |
| 0.75 | 0.14 |
| 0.875 | 0.07 |
| 1 | 0.0 |

1 D (debye) = $3.33564 \times 10^{-30}$ C.m

**Table S1**

### SI 3: Donor-Acceptor Coupling Energy

In our donor-acceptor photovoltaic model, the donor consists of two layers of WSeS with significant dipole moments arranged on top of each other, while the acceptor is a single layer of $WSe_2$ with zero dipole moment. The electric field due to the dipole moment in the donor layers was calculated by numerical methods for extended sheets of the donor layers. The induced polarization in the acceptor layer was calculated from the electric field from the donor layers, using Quantum ESPRESSO [3]. The donor-acceptor coupling energy ($\gamma_C$) and the corresponding exciton transfer rate was calculated from the dipole-induced dipole interaction energy between the WSeS layers and the WSe$_2$ layer. For the equilibrium donor-acceptor spacing of 3 Å, the coupling energy is 515 µeV which decreases with increasing donor-acceptor separation to a value of around 50 µeV for 10 Å separation. The donor layers exhibit dipole-dipole interaction between the individual layers.

$$U_{int} = -\frac{2\mu_1\mu_2}{4\pi\varepsilon_0 d^3}$$

where,
µ$_1$ and µ$_2$ – dipole moments of donor layers,
d – spacing between the donor layers.

The donor-acceptor interaction was a dipole-induced dipole interaction.

$$U_{int} = -\frac{\mu^2 \alpha}{(4\pi\varepsilon_0)^2 d^6}$$

where,
µ - dipole moment of donor,
α - polarizability of the acceptor,
d - donor acceptor separation.

Using the above formulas for isolated systems, we used numerical methods to obtain the interaction energy for extended sheets of the donor and acceptor materials.

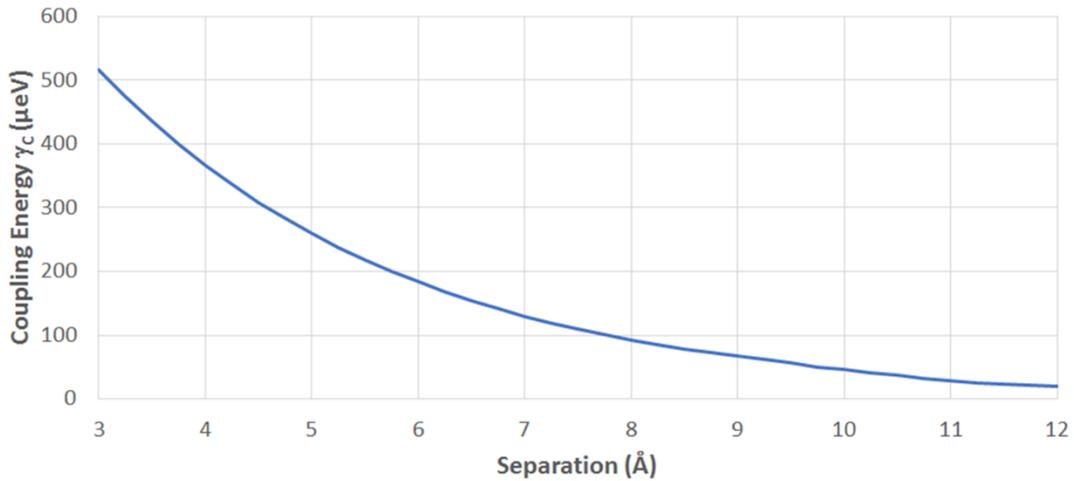

**Fig. S3.** Donor-Acceptor Coupling Energy $\gamma_c$

## SI 4: Current Enhancement due to Dark State Protection

The exciton transfer dynamics illustrated in Fig. 2(B) can be represented by the following Pauli master equations:

$$\begin{aligned}
\dot{\rho}_{ee} &= -\gamma_R \left[ \left(1 + n_h(E_e - E_g)\right)\rho_{ee} - n_h(E_e - E_g)\rho_{gg} \right] \\
&\quad -\gamma_C \left[ \left(1 + n_c(E_e - E_\alpha)\right)\rho_{ee} - n_c(E_e - E_\alpha)\rho_{\alpha\alpha} \right] \\
\dot{\rho}_{gg} &= \gamma_R \left[ \left(1 + n_h(E_e - E_g)\right)\rho_{ee} - n_h(E_e - E_g)\rho_{gg} \right] \\
&\quad +\gamma_C \left[ \left(1 + n_c(E_\beta - E_g)\right)\rho_{\beta\beta} - n_c(E_\beta - E_g)\rho_{gg} \right] \\
\dot{\rho}_{\alpha\alpha} &= \gamma_C \left[ \left(1 + n_c(E_e - E_\alpha)\right)\rho_{ee} - n_c(E_e - E_\alpha)\rho_{\alpha\alpha} \right] - \Gamma\rho_{\alpha\alpha} \\
\dot{\rho}_{\beta\beta} &= -\gamma_C \left[ \left(1 + n_c(E_\beta - E_g)\right)\rho_{\beta\beta} - n_c(E_\beta - E_g)\rho_{gg} \right] + \Gamma\rho_{\alpha\alpha}
\end{aligned} \quad (1)$$

and the dynamics illustrated in Fig. 2(C) can be represented by:

$$\begin{aligned}
\dot{\rho}_{bb} &= -\gamma_R \left[ \left(1 + n_h(E_b - E_g)\right)\rho_{bb} - n_h(E_b - E_g)\rho_{gg} \right] \\
&\quad -\gamma_{bd} \left[ \left(1 + n_c(E_b - E_d)\right)\rho_{bb} - n_c(E_b - E_d)\rho_{dd} \right] \\
\dot{\rho}_{dd} &= \gamma_{bd} \left[ \left(1 + n_c(E_b - E_d)\right)\rho_{bb} - n_c(E_b - E_d)\rho_{dd} \right] \\
&\quad -\gamma_C \left[ \left(1 + n_c(E_d - E_\alpha)\right)\rho_{dd} - n_c(E_d - E_\alpha)\rho_{\alpha\alpha} \right] \\
\dot{\rho}_{gg} &= \gamma_R \left[ \left(1 + n_h(E_b - E_g)\right)\rho_{bb} - n_h(E_b - E_g)\rho_{gg} \right] \\
&\quad +\gamma_C \left[ \left(1 + n_c(E_\beta - E_g)\right)\rho_{\beta\beta} - n_c(E_\beta - E_g)\rho_{gg} \right] \\
\dot{\rho}_{\alpha\alpha} &= \gamma_C \left[ \left(1 + n_c(E_d - E_\alpha)\right)\rho_{dd} - n_c(E_d - E_\alpha)\rho_{\alpha\alpha} \right] - \Gamma\rho_{\alpha\alpha} \\
\dot{\rho}_{\beta\beta} &= -\gamma_C \left[ \left(1 + n_c(E_\beta - E_g)\right)\rho_{\beta\beta} - n_c(E_\beta - E_g)\rho_{gg} \right] + \Gamma\rho_{\alpha\alpha}
\end{aligned} \quad (2)$$

where each $\rho_{xx}$ is the population in the $x$ state, the $n_h(E)$ is the optical distribution number defined by $n_h(E) = \left(e^{E/k_B T_h} - 1\right)^{-1}$ with $T_h = 5800K$ (temperature of the sun), and the $n_c(E)$ is the thermal distribution number defined by $n_c(E) = \left(e^{E/k_B T_c} - 1\right)^{-1}$ with $T_c = 300K$ (room temperature). The rates $\gamma_R$ and $\gamma_C$ are calculated in the main text. The thermal relaxation rate $\gamma_{bd}$ is usually very large compared to all the rates involved, and here we define it to be $\gamma_{bd} = 1000\gamma_R$ in accordance with previous studies [15-19]. The rate $\Gamma$ defines the output current as $I = e\Gamma\rho_{\alpha\alpha}$ where $e$ here is the elementary charge.